# Anisotropic hyperfine interactions in FeP studied by $^{57}$Fe Mössbauer spectroscopy and $^{31}$P NMR

Igor A. Presniakov,[1] Alexey V. Sobolev,[1] Andrey A. Gippius,[2,3]
Ivan V. Chernyavskii,[1] Martina Schaedler,[3] Norbert Buettgen,[3] Sergey A. Ibragimov,[1]
Igor V. Morozov,[1] and Andrei V. Shevelkov[1]

[1] Department of Chemistry, Lomonosov Moscow State University, 119991 Moscow, Russia

[2] Department of Physics, Lomonosov Moscow State University, 119991 Moscow, Russia

[3] Experimental Physics V, University of Augsburg, 86159 Augsburg, Germany

* corresponding author, Department of Chemistry, Lomonosov Moscow State University, Leninskie Gory 1-3, 119991 Moscow, Russia, alex@radio.chem.msu.ru

We report results of $^{57}$Fe Mössbauer and $^{31}$P NMR studies of a phosphide FeP powder sample performed in a wide temperature range including the point ($T_N \approx 120$ K) of magnetic phase transitions. The $^{57}$Fe Mössbauer spectra at low temperatures $T < T_N$ consist of very diffuse Zeeman pattern with line broadenings and sizeable spectral asymmetry. It was shown that the change of the observed spectral shape is consistent with the transition into the space modulated helicoidal magnetic structure. Analysis of the experimental spectra was carried out assuming the anisotropy of the magnetic hyperfine field $H_{hf}$ at the $^{57}$Fe nuclei when the $Fe^{3+}$ magnetic moment rotates with respect to the principal axis of the electric field gradient (EFG) tensor. The obtained large temperature independent anharmonicity parameter $m \approx 0.96$ of the helicoidal spin structure results from easy-axis anisotropy in the plane of the iron spin rotation. It was assumed that the very low maximal value of $H_{hf}(11K) \approx 36$ kOe and its high anisotropy $\Delta H_{anis}(11K) \approx 30$ kOe can be attributed to the stabilization of iron cations in the low-spin state ($S_{Fe} =1/2$). The $^{31}$P NMR measurements demonstrate an extremely broad linewidth reflecting the spatial distribution of the transferred internal magnetic fields of the $Fe^{3+}$ ions onto P sites in the magnetically ordered state.

**INTRODUCTION**

Binary compounds *MX* (where *M* is transition metal, *X* is pnictogen) of the MnP structure type are numerous, and many of them display magnetic properties that attract rapt attention. For instance, the first-order phase ferromagnetic transition in MnAs near room temperature gives rise to a pronounced magnetocaloric effect (MCE), with the change in magnetic entropy exceeding 20 J kg$^{-1}$ K$^{-1}$ at 293 K and 15 kOe, placing MnAs and its substituted variants among the most studied MCE materials [1, 2]. But arguably the most attractive property is heliomagnetism displayed by such compounds as MnP, FeP, and CrAs [3, 4]. The scientific interest into iron phosphide, which seems rather simple at the first glance, is largely associated with intrinsic complexity of its unusual helical magnetic structure, which details and generation mechanisms are still a matter of discussions. In general, there are two models proposed for the ordering of magnetic moments of iron cations; one of them assumes a unique type of crystallographic positions for Fe$^{3+}$, whereas the second model suggests two nonequivalent Fe1 and Fe2 positions with different magnetic moment values ($\mu_1$ = 0.37 $\mu_B$ and $\mu_2$ = 0.46 $\mu_B$) [5]. However, it should be noted that the neutron diffraction data applied to study the magnetic properties of FeP did not allow differentiating between these two models of magnetic ordering, suggesting that other techniques should be employed.

Early $^{57}$Fe Mössbauer works performed for FeP [6, 7] showed that all iron cations occupy a unique crystallographic positions within the FeP crystal structure in the paramagnetic temperature region ($T > T_N$), in accordance with the crystal data [8, 9]. However, identification and self-consistent analysis of the complex magnetic hyperfine spectra at $T < T_N$ caused serious difficulties [7] and left a number of questions remaining up to the date. Recently, a detailed Mössbauer study of isostructural arsenide FeAs was published [10], in which an original model of Mössbauer spectra fitting taking into account the modulated magnetic structure of FeAs ($T < T_N$ = 70 K [11]) was proposed, which is very similar to the magnetic ordering in FeP. However, despite the good description of all spectra measured in a wide temperature range ($T < T_N$), the model proposed in reference [11] has serious contradictions. In particular, one of the main conclusions drawn from the model fitting of Mössbauer spectra for FeAs was the existence of two nonequivalent positions of iron cations, which disagrees with the structural data [11]. Moreover, the authors of this paper did not explain the rather unusual profile of the spatial anisotropy of Fe1 and Fe2 magnetic moments, which requires independent experimental confirmation.

Therefore, it seems desirable to revisit the magnetic structure of FeP, employing $^{57}$Fe Mössbauer and $^{31}$P NMR spectroscopy in order (i) to determine how many crystal and magnetic iron-sites are distinguishable in the paramagnetic and magnetically ordered phases and (ii) to study the anisotropic shape of the magnetic hyperfine fields at $^{57}$Fe nuclei versus rotation of the iron moments in the (*ab*) plane of the orthorhombic unit cell. Besides, $^{31}$P NMR measurements will allow to answer the question whether the magnetic anisotropy affects the hyperfine magnetic fields at $^{31}$P nuclei transferred from the nearest magnetic $Fe^{3+}$ ions.

Recently [12], we proposed a new model for the fitting and analysis of complex hyperfine structure of $^{57}$Fe Mössbauer spectra for FeP recorded at $T < T_N$. It was shown the efficiency of this local method which provides important information related to unusual long-range magnetic ordering in the iron phosphide. This paper deals with the detailed analysis of the temperature dependencies of hyperfine parameters and their discussion in light of the peculiarities of the electronic and magnetic state of the iron cations in FeP. The results of a $^{57}$Fe Mössbauer study in a wide range of temperatures including the magnetic transition temperature ($T_N$) supplemented by first results of $^{31}$P NMR spectroscopy in iron phosphide FeP. Analysis of the experimental spectra was carried out assuming the anisotropy of the magnetic hyperfine field $H_{hf}$ at the $^{57}$Fe nuclei, when the $Fe^{3+}$ magnetic moment rotates with respect to the principal axis of the EFG tensor, and assuming the anisotropy of the magnetic hyperfine field $H_{hf}$ at the $^{57}$Fe nuclei.

**EXPERIMENTAL PART**

Synthesis was carried out by heating the stoichiometric mixture of Fe and P (red) powders in an evacuated quartz ampoule. The mixture was slowly (over 30 h) heated up to 1123 K and annealed at this temperature for 48 hours, after which the oven is put out. Powder X-ray diffraction analysis was performed on a Bruker D8 Advance diffractometer (Cu-K$_{\alpha 1}$ radiation, Ge-111 monochromator, reflection geometry) equipped with a LynxEye silicon strip detector. It revealed the obtained product to be pure FeP phase with the orthorhombic unit cell: $a$ = 5.203(1) Å, $b$ = 3.108(1) Å and $c$ = 5.802(1) Å, space group *Pnma*, which agrees perfectly with the literature data [5, 6]. All operations were carried out in a glovebox ($p$(H$_2$O, O$_2$) < 1 ppm) to prevent surface oxidation in a moist air.

Mössbauer spectra on $^{57}$Fe nucleus were obtained employing MS-1104Em spectrometer, operating in constant acceleration mode. Spectra were processed with SpectrRelax program [13]. All isomer shifts are given relative to the Mössbauer spectra of α-Fe at room temperature.

The $^{31}$P NMR measurements were performed both in the paramagnetic (at 155 K) and in the magnetically ordered state (at 1.55 K) utilizing a home-built phase coherent pulsed NMR spectrometer. NMR spectra were measured by sweeping the magnetic field at several fixed frequencies in the wide range of 18–120 MHz, the signal was obtained by integrating the spin-echo envelope in the time domain.

**RESULTS AND DISCUSSION**

**1. Mössbauer spectrum in paramagnetic state**

The Mössbauer spectrum of the $^{57}$Fe nucleus (Fig. 1 a) measured above the magnetic ordering temperature of FeP ($T_N \approx 120$ K) presents a single quadrupole doublet with narrow ($W = 0.31(1)$ mm/s) and symmetrical lines, thus emphasizing the uniformity of structural positions of iron atoms in the phosphide [12]. The value of the isomer shift $\delta_{300K} = 0.31(1)$ mm/s corresponds to iron atoms in a formal oxidation state of "+3" [14]. It is interesting that the observed $\delta$ value matches typical values of isomer shifts for high-spin cations $Fe^{3+}(d^5)$, octahedrally surrounded by oxygen [15], where, in contrast to phosphides, the Fe–O chemical bonds are considered to be almost entirely ionic. Apparently, this coincidence is due to the fact that the Fe←X transfer of the electron density, caused by a high degree of covalency in metal-pnictogen bonds ($X =$ P, As, Sb), induces a simultaneous increase in the 4$s$- and 3$d$-orbital populations, which affects the value of the isomer shift in the opposite direction [14, 15]. Mutual compensation of these two effects possibly renders the resulting isomer shift "less sensitive" to specific chemical bonds of iron cations in pnictides, in contrast to iron oxides, in which the isomer shift value allows estimating the symmetry of local surrounding and spin state of $Fe^{3+}$ ions [15].

High quadrupole splitting of the doublet, $\Delta_{300K} = 0.57(1)$ mm/s, means that strong electric field gradient (EFG) appears on the $^{57}$Fe nuclei. In the case of high-spin ions $Fe^{3+}(S_{Fe} = 5/2)$ with a spherically symmetric $3d^5$ ($^6S$) electron shell, the major contribution to the EFG must be associated with the distortion of the crystalline environment of the target nucleus. However, our calculations of the "lattice" contribution to the EFG tensor $\{V_{ij}^{lat}\}_{i,j=x,y,z}$ using the crystallographic data for FeP [5] showed that it is impossible to reach an agreement between the experimental and theoretical values under any physically reasonable values of effective charges of iron ($Z_{Fe}$) and phosphorus ($Z_P$) ions. Thus, the symmetry of the crystal environment of the iron cations in the FeP crystal structure cannot cause such high values of the EFG at $^{57}$Fe nuclei. This result indicates the necessity of considering the electronic contributions to the EFG that are related to different electronic populations of 3$d$-orbitals of iron [15] and the overlapping effects of filled 3$p$-orbital of $Fe^{3+}$ and valence 3($sp$) orbitals of phosphorus (*overlap distortion*) [17]. According to the $V^{lat}$ calculations, irrespective of the $Z_{Fe}$ and $Z_P$ values, the principal $V_{YY}$ axis for the four crystallographically equivalent Fei atoms in the orthorhombic unit cell lies along the $b$ axis (Fig. 1 b). The largest $V_{ZZ}$ component ($|V_{ZZ}| \geq |V_{YY}| \geq |V_{XX}|$) for all Fei atoms is located in the ($ac$) plane, but the angles ($\Theta_i$) which these components make with the $a$ axis are different for the four iron atoms: $\Theta_1 \approx 78°$, $\Theta_2 = \pi - \Theta_1$, $\Theta_3 = \pi + \Theta_1$, $\Theta_4 = \pi - \Theta_1$ (Fig. 1 b). Variation of the $Z_{Fe}$ and $Z_P$ charges leads only to a slight turn of the principal axes $V_{ZZ}$ and $V_{XX}$ in the plane (*ac*) of the FeP lattice.

## 2. Mössbauer spectra for magnetic ordering

Decreasing the temperature down to $T \approx T_N$ did not reveal any specific features in the monotonic temperature dependence of the hyperfine-interaction parameters ($\delta$, $\Delta$) that is usually observed for the $Fe^{3+}$ ions. At the same time, a complex Zeeman structure appears in the spectra upon moving into the low-temperature region $T < T_N$ (Fig. 3a), indicating the hyperfine magnetic fields $H_{hf}$ induced at $^{57}$Fe nuclei. Since the maximum value of the field $H_{hf}$ does not exceed 36 kOe even at the lowest temperature (11 K) and the quadrupole coupling constant reaches the value of $eQV_{ZZ} = |1.15(1)|$ mm/s (300 K), we used the full Hamiltonian of hyperfine interactions having the simplest form in the coordinate system of the principal axes of the EFG tensor [12]:

$$\hat{H}_{\mu Q} = \frac{eQV_{ZZ}}{4I(2I-1)}\left[3\hat{I}_Z^2 - \hat{I}^2 + \eta(\hat{I}_X^2 - \hat{I}_Y^2)\right] - g\mu_N H_{hf}\left[(\hat{I}_X \cos\varphi + \hat{I}_Y \sin\varphi)\sin\theta + \hat{I}_Z \cos\theta\right], \quad (1)$$

where $\hat{I}_X, \hat{I}_Y, \hat{I}_Z$ are the angular momentum operators of the $^{57}$Fe nucleus in its exited state; $g$ is nuclear $g$-factor; $\mu_N$ is the nuclear Bohr magneton, and $\eta = (V_{YY} - V_{XX})/V_{ZZ}$ is the parameter of asymmetry. The eigenvalues $\hat{H}_{\mu Q}$ depend not only on the hyperfine parameters of the system ($\delta$, $eQV_{ZZ}$, $H_{hf}$, $\eta$), but also on spherical angles ($\theta$, $\varphi$) that determine the orientation of the hyperfine field $H_{hf}$ in the system defined by the principal axes of the EFG tensor ($X,Y,Z$) (Fig. 2 a). The eigenvalues of such Hamiltonian depend on a number of parameters ($\delta$, $eQV_{ZZ}$, $H_{hf}$, $\eta$, $\theta$, $\varphi$), many of them are correlated, *i.e.*, cannot be determined independently from Mössbauer spectra, but only in certain combinations. However, despite the large number of variable independent hyperfine parameters we could not adequately describe the experimental spectra profiles at $T < T_N$ according to this simple model. In Fig. 3a (blue shaded area), we present an example of the best possible description of the experimental spectrum at $T = 11$ K, which accounts for the main features, but can not reconstruct the profile completely. It is important to note, that in the case of FeP lattice there are four crystallographically equivalent Fei sites with different directions of local principal EFG axes (Fig. 1 b), and, as consequence, different $\theta_i$ and $\varphi_i$ values. However, due to the symmetry of the complete hyperfine Hamiltonian (1), if the magnetic structure of FeP would be collinear, then for these four values of $\theta_i$ and $\varphi_i$, angles the positions and the relative intensities of the nuclear Zeeman patterns will be the same irrespective of the value of $\eta$ [6].

## 3. Helicoidal structure

As long as the model with a unique iron site failed to completely describe the experimental spectrum, in the next step of the analysis at $T < T_N$ we used a more sophisticated model by taking into account the features associated with an incommensurate planar elliptical structure for FeP, which is schematized in Fig. 2b. Following the helicoidal structure proposed in [5], the magnetic moments are parallel within the planes 1, 2, 3, 4 normal to the $c$ axis. As the plane changes 1-3 or 4-2, the direction of the spin rotates by $\sim 36^0$ (the difference of the angles between the spin directions of two Fe atoms separated by $0.4c$) with the relative angle $176^0$ between adjacent planes separated by $0.1c$ (Fig. 2 b). Such rotation between adjacent moments along the $c$ axis implies the presence of a quasi-continuum of possible orientations of iron moments $\mu_{Fe}$ lying in the ($ab$) plane. The angle $\vartheta$, which gives the orientation of the hyperfine field $H_{hf}$ in the ($ab$) plane, is varying continuously between 0 and $2\pi$. We found the expressions $\theta(\vartheta, \Theta)$ and $\varphi(\vartheta, \Theta)$ connected the spherical angles in Hamiltonian $\hat{H}_{\mu Q}$ (1) with the rotation angle ($\vartheta$) of the magnetic moments of iron ions and also with the angle $\Theta$ that the principle $V_{ZZ}$ component makes with the normal (the $c$-axis) to the helical plane (Fig. 2 a):

$$\cos\theta = \sin\Theta\cos(\vartheta) \qquad (2a)$$

$$\text{tg}\varphi = \text{tg}(\vartheta)/\cos\Theta \qquad (2b)$$

The values $\vartheta_i(z)$ depend on the coordinates of iron atoms ($z$) along the helix propagation (Fig. 2b), in particular, for the harmonic approximation $\vartheta(z) = kz + \tau$ (where $k$ is the wave vector and $\tau$ is the phase shift). Our model assumes the linear relation of the magnetic moment $\mu_{Fe}$ and the hyperfine magnetic field $H_{hf}$ at $^{57}$Fe nuclei: $H_{hf} = \alpha\mu_{Fe}$ (neglecting the influence of magnetic neighbors – *supertransferred hyperfine field*, $H_{STHF}$). To account for the possible spatial anisotropy of the magnetic moment $\mu_{Fe}(\vartheta) = a\mu_a\cos\vartheta + b\mu_b\sin\vartheta$ or the anisotropy of the magnetic hyperfine tensor ($\tilde{A}$), both of which leads to anisotropy of the $H_{hf}$ field, we used the angular dependence $H_{hf}(\vartheta) = H_a\cos^2\vartheta + H_b\sin^2\vartheta$, where $H_a$ and $H_b$ are the magnitudes of $H_{hf}$ when the magnetic moment of $Fe^{3+}$ is directed along the $a$ and $b$ axis, respectively. Finally, we used Jacobian elliptic function to describe the anharmonicity (bunching) of spatial distribution of $Fe^{3+}$ magnetic moments [18]:

$$\cos\vartheta(z) = \text{sn}[(\pm 4K(m)/\lambda)z, m], \qquad (3)$$

where K($m$) is the complete elliptic integral of the first kind, $m$ the anharmonicity parameter, $\lambda$ the helical period. Depending on the parameter $m$, the spin modulation changes from purely circular helicoid ($m = 0$ for $K_u = 0$) to a square wave ($m \to 1$ for $|K_u| \gg 1$), in which spins bunch along the modulation propagation $z \| c$ direction. The use of this model allowed to satisfactorily describe the entire series of experimental spectra measured in the magnetic ordering temperature range 11 K $\leq T < T_N$ (Fig. 4).

## 4. Anisotropy and anharmonicity

It is possible to make a number of observations and comments based on the acquired data. First of all, one should pay attention to a very small value of the maximal value of magnetic field $H_a$(11 K) = 35.9(1) kOe. It is usually assumed that the Fermi contact interaction with *s*-electron is the dominant hyperfine coupling, implying that the Fe moment is proportional to the hyperfine field via the relation $H_{hf} = \alpha\mu_{Fe}$, where the value of the proportionality constant $\alpha$ is compound specific. In our experiments, the average $\alpha$ value was obtained from $H_{hf}$ and $\mu_{Fe}$ determined from Mössbauer and neutron diffraction study of isostructural pnictides $AFe_2As_2$ ($A$ = Ca, Sr, Ba) [19]. Using thus obtained the average ratio $\alpha \equiv H_{hf}/\mu_{Fe} \approx 92$ kOe/$\mu_B$, we estimated possible value of the magnetic moment to be $\mu_{Fe} \approx 0.39$ $\mu_B$ which appear to be very close to the values of 0.37 – 0.46 $\mu_B$ deduced from neutron diffraction study of FeP [5]. It should be noted that despite the small difference in values of $\mu_{Fe}$ defined by different methods, all of them clearly point at a significant decrease of the resultant spin of iron compared to $S_{Fe} = 5/2$ ($\mu_{Fe} = 5$ $\mu_B$) for high-spin cations $Fe^{3+}(d^5)$. Such a reduction of $S_{Fe}$ cannot be connected to the effects of strong covalency of the Fe-P bonds in (FeP$_6$) polyhedra of iron phosphide.

Our model suggests that the "spatial anisotropy of hyperfine field" refers to the angular dependency of the $H_{hf}(\vartheta)$ in the (ab) plane of the orthorhombic unit cell or $H_{hf}(\theta_\vartheta, \varphi_\vartheta)$ field value in the system defined by principal axes of the EFG tensor (Fig. 2a). The strong spatial anisotropy of hyperfine magnetic field $H_{hf}$, particularly at $T$ = 11 K, $H_a$ = 35.9(1) kOe and $H_b$ = 5.6(1) kOe, is related to the spatial anisotropy of magnetic moment values $\mu_{Fe}(\vartheta)$ [5] or hyperfine magnetic interaction constant $\tilde{A}(\vartheta)$ (generally, a tensor quantity) included in the expression $\boldsymbol{H}_{hf} = \tilde{A}\cdot\boldsymbol{\mu}_{Fe}$ [20]. Figure 5 represents an experimental polar diagrams $\boldsymbol{H}_{hf}(\vartheta)$ at different temperatures. As can be seen from the figure the difference ($H_a$ - $H_b$) $\equiv \Delta H_{anis}$ decreases monotonically with increasing temperature (Fig. 6 a). The anisotropy of $H_{hf}$ usually associates with the dipole contribution $H_{dip}$ due to the asymmetry of the local magnetic surrounding, but in our case it cannot significantly influence $H_{hf}$ because of the low $\mu_{Fe}$ value. In our case, the most probable cause of the anisotropy $H_{hf}$ is related with the anisotropy of the magnetic moment $\mu_{Fe}(\vartheta)$ as it rotates in the (ab) plane. This conclusion is consistent with the results of the FeP neutron diffraction study [5], according to which the anisotropy of magnetic moment is $\Delta\mu_{anis} \approx 0.34\mu_B$ ($T$ = 4.6 K); that should lead to the value $\Delta H_{anis} = \alpha\Delta\mu_{anis} \approx 31.3$ kOe that is almost completely coincides with the experimental value 30.3(2) kOe. In addition, we can not exclude the anisotropic part of the magnetic hyperfine interaction constant $\tilde{A}$ as the probable cause for the observed anisotropy $H_{hf}$ responsible for the mechanisms of formation of the magnetic hyperfine structure [20]. One should also consider the orbital contribution to $\tilde{A}$ that is strongly dependent on the iron spin direction towards the crystal axes.

Within the above model, the best description of all spectra measured at 11 K ≤ $T$ ≤ 120 K (Fig. 4) may only be attained with sufficiently high values of the anharmonicity parameter $m$ that remains almost unchanged ($m \approx 0.96$) in the entire temperature range ($T < 0.87 \cdot T_N$) of magnetic order (Fig. 6 a). The anharmonicity is related to the constant ($K_u$) of uniaxial magnetocrystalline anisotropy [18] that contains two main contributions: $K_u = K_D + K_{SI}$ [15], where $K_D$ is a dipole contribution connected to dipole-dipole interactions of magnetic moments $\mu_{Fe}$, which depend on their orientation in a crystal, and $K_{SI}$ is the single-ion anisotropy determined by the effects of the spin-orbital interaction [21]. In the case of FeP, the $K_D$ contribution should not play any significant role due to the very low values of magnetic moment $\mu_{Fe}(\vartheta)$. The manifestation of single-ion anisotropy $K_{SI}$ by the high-spin cations $Fe^{3+}$ is possible only with consideration of excited electronic states with a nonzero orbital angular moment [15].

## 5. First-order magnetic phase transition

To get an insight into the nature of the magnetic transition, the temperature dependence of the average hyperfine magnetic field $<H_{hf}(T)> = H_b(T) + \{H_a(T) - H_b(T)\}<sn^2\vartheta>$ (Fig. 6 b) was analyzed using the Bean-Rodbell (*B-R*) model [22], where the exchange magnetic interactions are considered to be a sufficiently strong function of the lattice spacing and compressibility. In the *B-R* model [22], the free-energy ($F$) without external field is expanded in terms of the one order parameter σ:

$$F(\sigma) \propto (T - T_N)\sigma^2 + 1/6(T - \zeta T_N)\sigma^4 + D\sigma^6, \quad (4)$$

where $\sigma = M_{Fe}(T)/M_{Fe}(0) \approx <H_{hf}(T)>/<H_{hf}(0)>$ is the reduced sublattice magnetization or the hyperfine magnetic field at $^{57}$Fe; $D > 0$, and the sign of the fourth-order term prefactor ($T - \zeta T_N$) is determined by the $\zeta$ fitting parameter, which in turn involves the magneto-structural coupling coefficient and elastic modulus. The value of this parameter controls the order of the magnetic phase transition; for an ideal second-order phase transition $\zeta = 0$ and for a first-order phase transition $\zeta > 1$ (the $\zeta$ values between 0 and 1 correspond to an intermediate-order transitions). According to the *B-R* model [22, 23], the reduced magnetization of the iron sublattice can be expressed as:

$$\sigma(T) = B_S\left[\frac{3S}{S+1}\frac{\sigma(T)}{\tau}\left(1 + \frac{3}{5}\frac{(2S+1)^4 - 1}{2(S+1)^3 S}\zeta\sigma(T)\right)\right]. \quad (5)$$

The fitted $\zeta = 1.16 \pm 0.08$ value for the FeP (taken for the low spin $Fe^{3+}$ ions $S = \frac{1}{2}$, see below for explanation) leads to the negative fourth-order term near the transition temperature that creates an energy barrier in the free energy $F(\sigma)$, indicating a first-order transition. Physical meaning of our results can be understood within the phenomenological Landau theory of the magnetic phase transitions in a compressible lattice [24]. In this theory, the free-energy density is expanded in terms of two order parameters ($\sigma$, $u$):

$$F(\sigma, u) \propto A\sigma^2 + B\sigma^4 + bu^2/2 + \lambda\sigma^2 u, \qquad (6)$$

where the third term is the elastic energy, $b$ is the bulk modulus (which is proportional to the inverse compressibility), and the last term is the coupling of order parameter $\sigma$ with the lattice deformation $u$ (here $\lambda$ is a magnetoelastic constant) [24]. Minimization of the total energy (6) with respect to $u$ gives a renormalized free-energy density $F(\sigma)$ with the renormalized coefficient $\propto (B - \lambda^2/2b)$ of the $\sigma^4$ term, the sign of which coincides with $(T - \zeta T_N)$ in Eq. (4) (negative for $\zeta > 1$). If the coupling to the lattice $\lambda$ is sufficiently strong or if the lattice compressibility is large, the fourth-order term in (4) becomes negative, setting the first order transition. The fitting parameter $\zeta \propto \lambda^2/T_0 b$ in Eq (4) involves the magnetoelastic coupling and the bulk modulus [24], so its value also predicts the strength of the magnetoelastic coupling and controls the order of the magnetic phase transition.

Therefore, according to the our data, FeP demonstrates strong coupling magnetic ordering to the lattice deformation leading to the first order magnetic phase transition. Such transitions have been observed in many compounds, for example, MnAs [1], $MnFeP_{1-x}As_x$ [25], NaFeAs [26], FeAs [10], and FeS [27]. Various mechanisms have been examined in order to interpret the first-order magnetocrystalline transition, such as exchange magnetostriction [27, 28], biquadratic exchange [29] or spin-orbital coupling of magnetization in the presence of orbital degeneracy (Jahn-Teller distortions) [30] and so on. However, due to the lack of sufficient experimental data, the authors were unable to reach a common final decision.

### 6. Low-spin model for $Fe^{3+}$

Thus, the above Mössbauer data suggest that the orbital angular moment contribution should be taken into account while analyzing the basic electronic state of the $Fe^{3+}$ cations in iron phosphide FeP. High-spin $Fe^{3+}$ ions in the distorted octahedral environment of phosphorus anions are characterized by the electronic term $^6A_{1g}$, which requires orbital moment to be completely frozen. Consideration of $<L> \neq 0$ is possible only in the second order of perturbation theory [24] that could hardly explain the low value of hyperfine magnetic field $H_{hf}$, the high value of anisotropy $\Delta H_{anis}$ and strongly pronounced anharmonicity ($m \to 1$) of magnetic helicoid observed in our experiments. A possible reason for all these results is the stabilization of iron cations in a low-spin state (LS). Examples of the high-spin to low-spin transition in isostructural compounds were observed for FeAs and P-doped MnAs at 77 and 232 K, respectively [9, 31]. In a strong crystal field of octahedral symmetry the ground state of the ferric $Fe^{3+}$ ions is $^2T_{2g}$ ($t_{2g}^5$, $S_{Fe} = 1/2$). Considering the relation between holes and electrons, we have to take the $t_{2g}^5$ electron configuration as a $t_{2g}^1$ hole configuration. If the local symmetry is exactly octahedral, then the $^2T_{2g}$ state is sixfold degenerate. However, under the influence of axial tetragonal or trigonal distortions together with spin-orbit coupling, the sixfold degeneracy is removed (Fig. 7 a). According to our estimations, the negative sign of $eQV_{ZZ}$ ($< 0$), which is observed at $T \ll T_N$, is compatible with a crystal field with predominant tetragonal or very large ($\varepsilon/\Delta > 0.4$, see Fig. 7a) and physically unreasonable trigonal distortion (small trigonal character gives a positive sign for the quadrupole interaction). Thus, although there is no direct experimental evidence for the presence of a fourfold local symmetry axis at the $Fe^{3+}$ positions in the FeP lattice, the sign of the measured EFG tensor shows that the iron ions experience a quasi-tetragonal crystal field.

The hyperfine interactions in low-spin iron ions are considerably modified as compared with those of high-spin iron ions. For instance, despite the fact that the axial and rhombic distortions of the ($Fe^{3+}P_6$) polyhedra causes the splitting of sub-levels $t_{2g} \rightarrow a_{1g} + e_g$ (Fig. 7a), a small difference in energies of the generated terms $^2A_{1g}$ and $^2E_g$ as well as the spin-orbital interaction ($\lambda LS$) will lead to a partial "unfreezing" of the orbital moment $<L_i>$

$$\langle L_i \rangle = \lambda \Lambda^{ij} S_j = \lambda \sum_n \frac{\langle 0|\hat{L}_i|n\rangle \langle n|\hat{L}_j|0\rangle}{E_n - E_0} \cdot S_j, \qquad (7)$$

where $\lambda$ is the spin-orbital interaction constant; $\hat{L}_i$ is the orbital angular momentum operator; $<0|$ and $|n>$ are wave vectors of ground and exited electronic states; $E_0$ and $E_n$ are the energies of the corresponding states; $\Lambda^{ij}$ are the ($\tilde{\Lambda}$) tensor components that define the anisotropy of the orbital moment. As Fermi ($H_F$) and orbital $H_L = \lambda \tilde{\Lambda} \cdot S \mu_B <r^{-3}>_{3d}$ contributions to the hyperfine magnetic field have opposite signs, their partial canceling may lead to a significant decrease of the $H_{hf}$ value. In addition, since the value of the orbital moment $<L_i>$ is mostly anisotropic ($\Lambda^{xx} \approx \Lambda^{yy} \neq \Lambda^{zz}$), the $H_L$ contribution induced by this moment must be anisotropic as well. Finally, "mixing" the main electronic state of low-spin cations $Fe^{III}$ and the term $^2E_g$ leads to the Ising-like anisotropy [21], which could manifest itself as a high anisotropy level of spiral magnetic structures.

The above assumption about low-spin state of $Fe^{3+}$ ions in FeP provides good explanations for the main features of the experimental temperature dependence of a quadrupole coupling constant $eQV_{ZZ}$ at $T < T_N$ (Fig. 7b). If in high-spin $Fe^{3+}$($t_{2g}^3 e_g^2$) compounds the EFG acting on the iron nuclei is of lattice origin ($V^{lat}$), then in low-spin $Fe^{3+}(t_{2g}^5)$ state it is mainly produced by $3d^5$-electrons of iron ions. The temperature dependent electron population on $t_{2g}$ sublevels produces a relatively strong temperature dependent electronic contribution ($V^{el}$) to the EFG [16, 20]. Taking into account the temperature dependent probabilities of a hole occupying the $^2A_{1g}$ and $^2E_g$ levels, the temperature dependence of the electronic EFG can be described as [16]

$$V_{ZZ}(T) = V_{ZZ}(0)[(1 - exp(-\varepsilon/k_BT))/(1 + 2exp(-\varepsilon/k_BT))], \qquad (8)$$

From the thermal variation of $eQV_{ZZ}$ we can give an evaluation of the crystal field splitting ($\varepsilon$) between the two lowest orbitals of the low-spin $Fe^{3+}$ ions (Fig. 7a) in the FeP structure. The EFG is a decreasing function of increasing temperatures; this effect, which arises from the fact that the EFG is a thermal average between the values corresponding to the lowest orbital doublet and singlet. The axial case (only one free parameter, $\varepsilon$) proved to be the best in all fitting procedures. Using the data of Fig. 7b and the simple formula (8) one gets $\varepsilon = 2.61$ kJmol$^{-1}$K$^{-1}$. As can be seen (Fig. 7b), the agreement of calculated curve with experimental results is quite satisfactory. In the paramagnetic range $T > T_N$, the temperature dependence of the quadrupole interactions has been described using a semi-empirical relation $V_{ZZ}(T) = A(1 - B \cdot T^{3/2})$, where $A \equiv V_{ZZ}(T_N)$ is the main EFG tensor component at $T \approx T_N$, corresponding to the $\Delta_{118K} = 0.61(1)$ mm/s value of quadrupole splitting, and B is a constant with positive value $1.17(6) \cdot 10^{-5}$ K$^{2/3}$. In this temperature range ($T > T_N$), the observed $eQV_{ZZ}(T)$ dependence is mainly due to the temperature variation of the lattice $V^{lat}(T)$ contribution to the EFG [20].

### 7. $^{31}$P NMR spectra

The remaining question is whether the magnetic anisotropy affects the hyperfine magnetic fields at $^{31}$P nuclei transferred from the nearest magnetic $Fe^{3+}$ ions. Fortunately, apart from the most popular Mössbauer nuclei, $^{57}$Fe, the crystal structure of FeP contains also very good NMR nuclei $^{31}$P ($I = 1/2$, $\gamma/2\pi = 17.235$ MHz/T). This enables us to probe the hyperfine interactions on the phosphorus site in FeP by $^{31}$P NMR spectroscopy.

The field-sweep $^{31}$P NMR spectrum of the FeP powder sample measured at fixed frequency of 80 MHz in the paramagnetic state at 155 K is presented in Fig. 8 a. The line is very narrow (FWHM ~ $6 \cdot 10^{-2}$ kOe) with the peak situated almost at the diamagnetic Larmor field $H_{Lar} = 46.42$ kOe.

With decreasing temperature below $T_N$ the $^{31}$P spectra change dramatically. The spectrum measured at lowest temperature of 1.55 K is presented in the Fig. 8b. First of all, the spectrum is now extremely broad with FWHM ~ 11.6(4) kOe and width at the line basement ~ 16.4(4) kOe which is more than two orders of magnitude higher than the FWHM in the paramagnetic state.

This result unambiguously indicates that in the magnetically ordered state the effective magnetic field on $^{31}$P nuclei is strongly affected by the hyperfine field transferred from $Fe^{3+}$ cations. The spatial distribution of these transferred fields with respect to external magnetic field is reflected in the characteristic trapezoidal $^{31}$P line shape.

The trapezoidal distribution of the resonance fields $H_{res}$ in an antiferromagnet can be modeled based on the superposition of the internal magnetic field $H_{int}$ and the externally applied magnetic field $H$ [32, 33]:

$$H_{res} = \omega_0/\gamma = \left|\vec{H} + \vec{H}_{int}\right|, \tag{9}$$

or, equivalently,

$$H_{res}^2 = H^2 + H_{int}^2 + 2H \cdot H_{int} \cdot \cos\alpha, \tag{10}$$

where $\alpha$ is the angle between $H$ and $H_{int}$. In a powder sample, the direction of $H$ is randomly distributed with respect to $H_{int}$, and the probability for an angle between $H$ and $H_{int}$ to be in an interval between $\alpha$ and $\alpha + \Delta\alpha$ is proportional to the solid angle $\Delta\Omega \sim \sin\alpha \cdot \Delta\alpha$. The number of nuclei $\Delta N = f(H)\Delta H$ which probe a resonance field between $H$ and $H + \Delta H$ is proportional to the solid angle $\Delta\Omega$ or $f(H)\Delta H \sim \sin\alpha \cdot \Delta\alpha$, giving the NMR line shape $f(H)$:

$$f(H) \propto \sin\alpha \frac{d\alpha}{dH}. \tag{11}$$

The latter term $d\alpha/dH$ is given by the differentiation of Eq. (10) using the relation:

$$\frac{d\alpha}{d\cos\alpha} \cdot \frac{d\cos\alpha}{dH} = -\frac{1}{\sin\alpha} \cdot \frac{H^2 - H_{int}^2 + H_{res}^2}{H_{int} \cdot H^2}, \tag{12}$$

This result, inserted into Eq. (11), gives $f(H)$ which represents the number of nuclei which are in resonance at a certain value of the applied magnetic field $H$. As different values of $H_{res}$ are only obtained for angles $0 < \alpha < \pi$, the function $f(H)$ is valid only in the region $H_{res} - H_{int} < H < H_{res} + H_{int}$ with edge singularities at $H_{res} \pm H_{int}$, whereas it is zero out of this range (cf. Eq. (12)). For the simulation of the intensity $I(H)$ of real powder spectra a convolution with a Gaussian distribution of internal fields has to be accomplished:

$$I(H) \propto \int_{H_{res}-H_{int}}^{H_{res}+H_{int}} \frac{H^2 - H_{int}^2 + H_{res}^2}{H_{int} \cdot H^2} \cdot \frac{1}{\Delta H_{int}} \cdot \exp\left[-\frac{1}{2} \cdot \frac{(t-H)^2}{\Delta H_{int}^2}\right] dt. \tag{13}$$

The red solid line in Figure 8 b is a result of simulation of the experimental spectrum using Eq. (13) with the following parameters: $H_{res}$ = 34.1 kOe, $H_{int}$ = 6.8 kOe, $\Delta H_{int}$ = 0.7 kOe. As seen from this figure, the coincidence with the experimental $^{31}$P spectrum is quite satisfactory. Some deviation is expected since in the above analysis we assumed an isotropic distribution of the hyperfine field orientation with respect to the external magnetic field. Accounting for the anisotropy and anharmonicity effects will lead to characteristic "bumping" of the spectra (see, for instance, [34]).

# CONCLUSIONS

The results obtained in the present study have shown the effectiveness of the suggested approach to the complex hyperfine magnetic structure analysis of the experimental spectra of iron phosphide FeP measured over a wide temperature range. It is established that the abnormally low value of hyperfine magnetic field $H_{hf}(11K) \approx 36$ kOe and its high anisotropy $\Delta H(11K) \approx 30$ kOe can be related to the stabilization of iron cations in the low-spin state ($S_{Fe} = 1/2$). This assumption is consistent with the high value of the anharmonicity parameter of helical structure obtained from the spectra of interest. Our preliminary $^{31}$P NMR measurements on FeP demonstrate extremely broad spatial distribution of the transferred internal magnetic fields on P sites in the magnetically ordered state.

**Acknowledgments**

This work is supported in part by the Russian Science Foundation, grant # 14-13-00089 and Russian Foundation for Basic Research, grant # 15-03-99628

**FIGURES**

**Figure 1.** (*Color online*) (a) $^{57}$Fe Mössbauer spectrum of FeP recorded at $T = 300$ K ($T \gg T_N$). The solid line is the result of simulation of the experimental spectra as described in the text. (b) Directions of the principal EFG axes for the four equivalent Fei sites in unit cell of FeP.

**Figure 2.** (*Color online*) (a) The local magnetic structure (for one iron site) of FeP. The angle $\vartheta$ gives the orientation of the hyperfine field $H_{hf}$ in the (*ab*) plane, varying continuously between 0 and $2\pi$; the polar angle $\Theta$ gives the orientation of the principal $V_{ZZ}$ component in the (*ac*) plane (when the *c* axis is normal to the plane of the $H_{hf}(\vartheta)$ ellipse); the polar ($\theta$) and azimutal ($\varphi$)

angles give orientation of the $H_{hf}$ field in the coordinate system of the principal axes of the EFG tensor. (b) Schematic view of the double helical spin structure and its *ab* projections proposed in [5] from neutron diffraction study. The orange, blue, white and dark colored circles illustrate the iron atoms lying in the different (*ab*) planes (the spins are parallel within each plane normal to the *c* axis).

**Figure 3**. (*Color online*) (a) $^{57}$Fe Mössbauer spectrum at 11K fitted using a modulation of the hyperfine interactions as the $Fe^{3+}$ magnetic moment rotates with respect to the principal axis of the EFG tensor, and the anisotropy of the magnetic hyperfine interactions at the $Fe^{3+}$ sites. (b) The hyperfine field distribution $p(H_{hf})$ resulting from simulation of the spectra (red area). For comparison, the figure shows the distribution $p(H_{hf})$ for a harmonic helicoidal magnetic structure (blue area). Insert: modulation of the projection ($H_a$) of the $H_{hf}$ field on the *a* axis along the direction *q*, derived from the Mössbauer spectrum (red line); for comparison, blue line gives the theoretical curve for harmonic (sin-modulated) spin structure.

**Figure 4**. (*Color online*) $^{57}$Fe Mössbauer spectra (experimental hollow dots) of FeP recorded at the indicated temperatures. The solid red lines are simulation of the experimental spectra as described in the text.

**Figure 5.** (*Color online*) Polar diagram demonstrating anisotropy of the hyperfine magnetic fields at $^{57}$Fe nuclei in FeP for selected temperatures; the symbol $H_a$ denotes hyperfine field component on iron along the *a*-axis, while $H_b$ stands for the iron hyperfine field component along the *b*-axis)

**Figure 6.** (*Color online*) (a) Temperature dependences of anharmonicity parameter (*m*) and difference $\Delta H_{anis} = (H_a - H_b)$ extracted from least-squares fits of the Mössbauer spectra (b) Average hyperfine fields <$H_{hf}$> plotted versus temperature; fitting to the experimental data (red solid line) with the Bean-Rodbell model (inset: temperature dependence of larger ($H_a$) and smaller ($H_b$) components).

**Figure 7.** (*Color online*) (a) Low-spin $Fe^{3+}$ orbital level scheme in the trigonal crystal field approximation; (b) Quadrupole coupling constant $eQV_{ZZ}$ plotted versus temperature; fitting to the experimental data (blue solid line) with crystal field model (the region I – $T < T_N$) and using semi-empirical relation (see text) in the region II – $T > T_N$

**Figure 8**. (*Color online*) (a) $^{31}$P NMR spectrum of FeP measured in the paramagnetic state at 155 K; (b) $^{31}$P NMR spectrum of FeP measured in the magnetically ordered state at 1.55 K. (the red solid line is the simulation, see text).

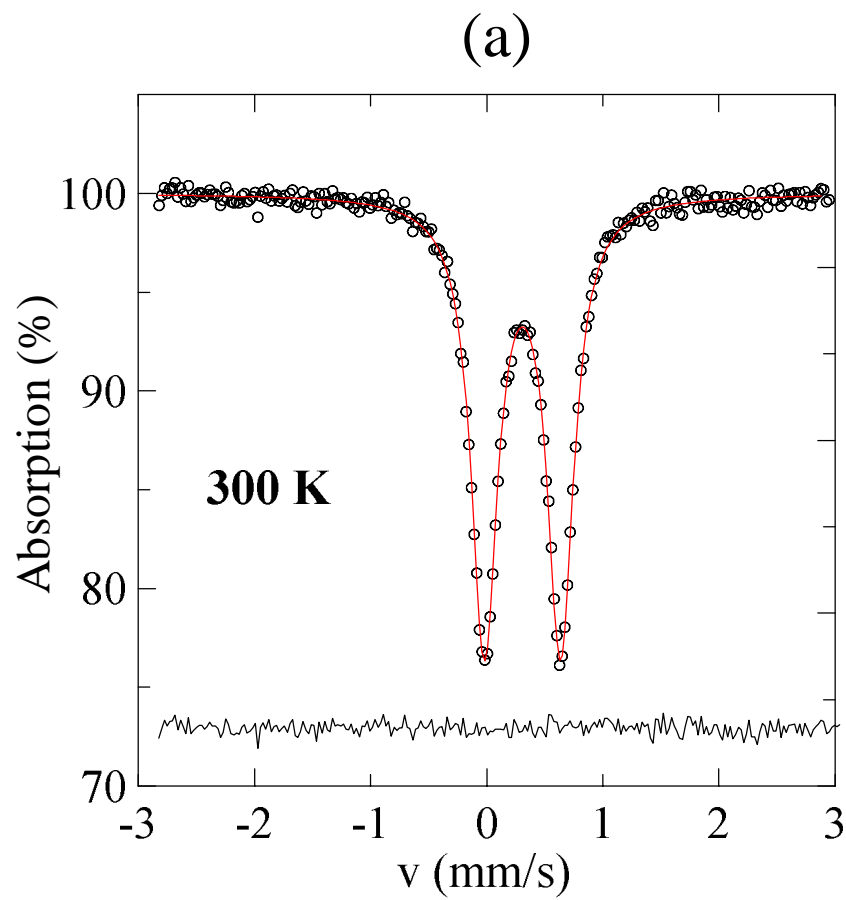

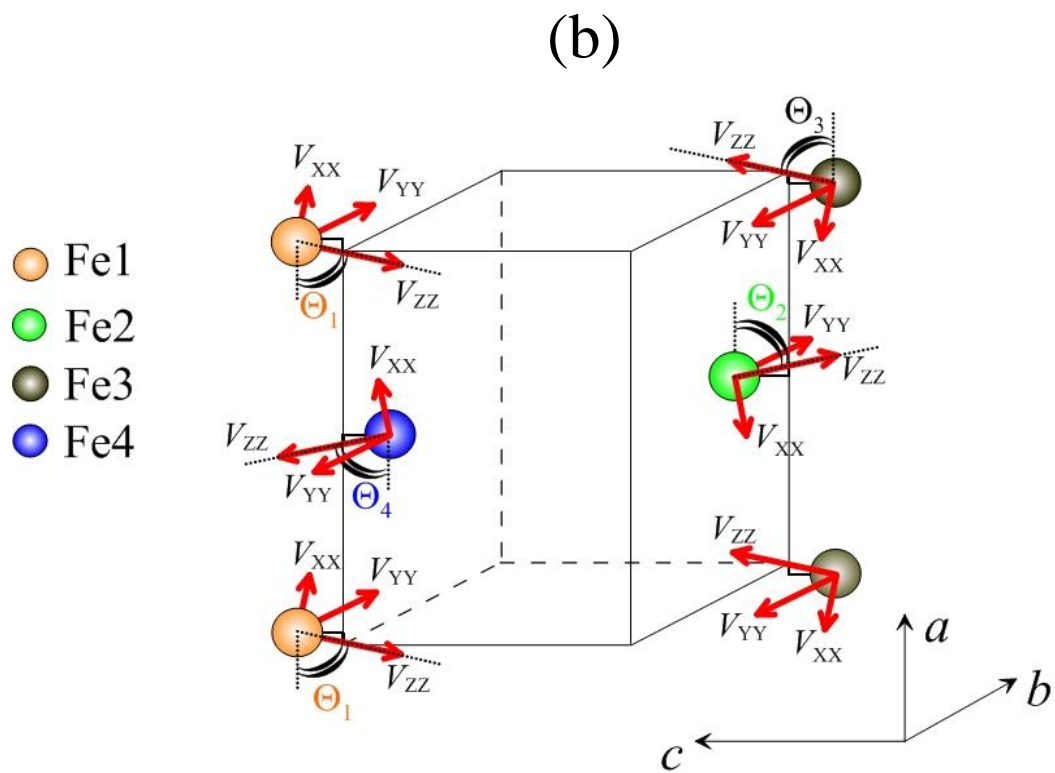

**Fig. 1**

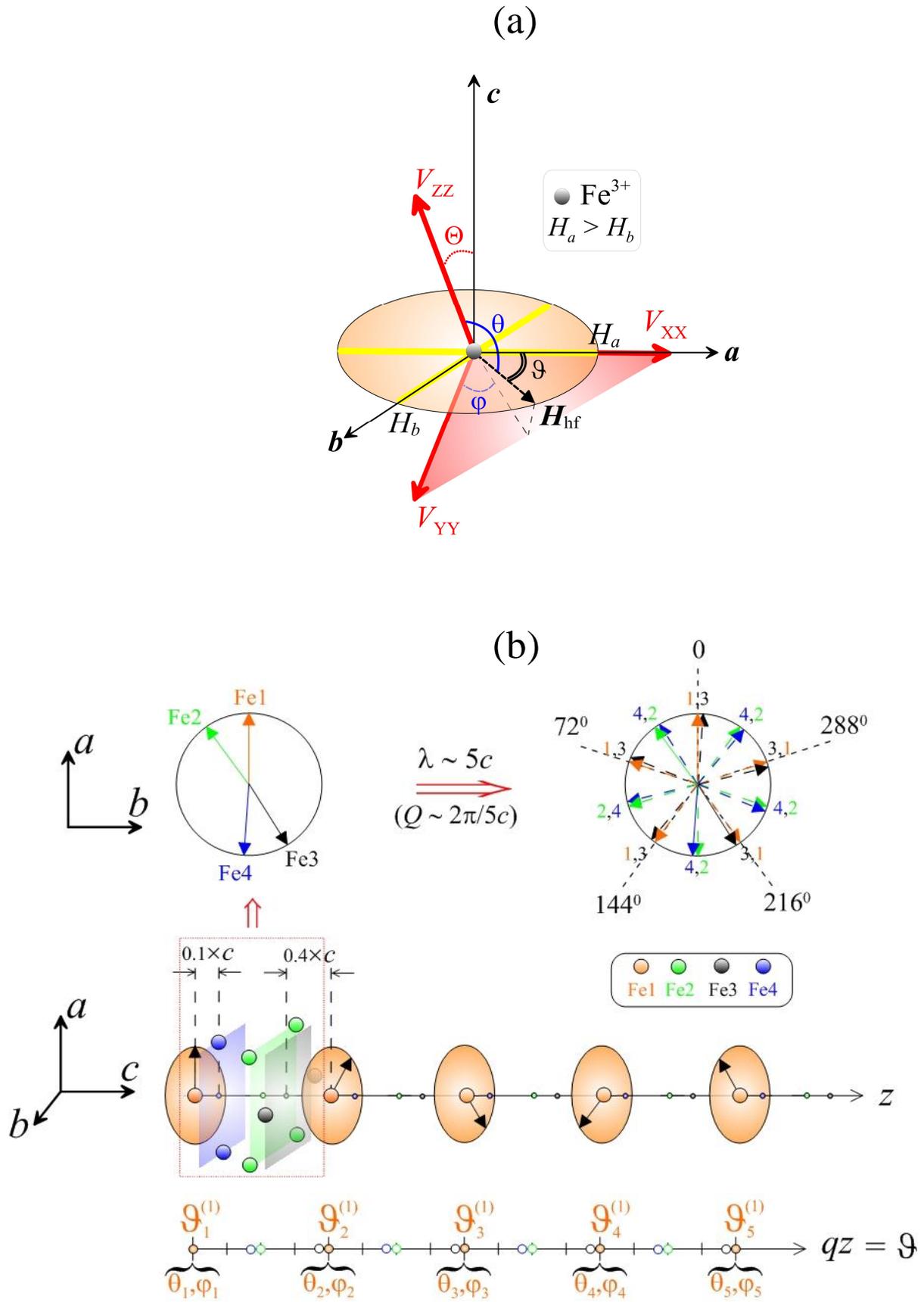

**Fig. 2**

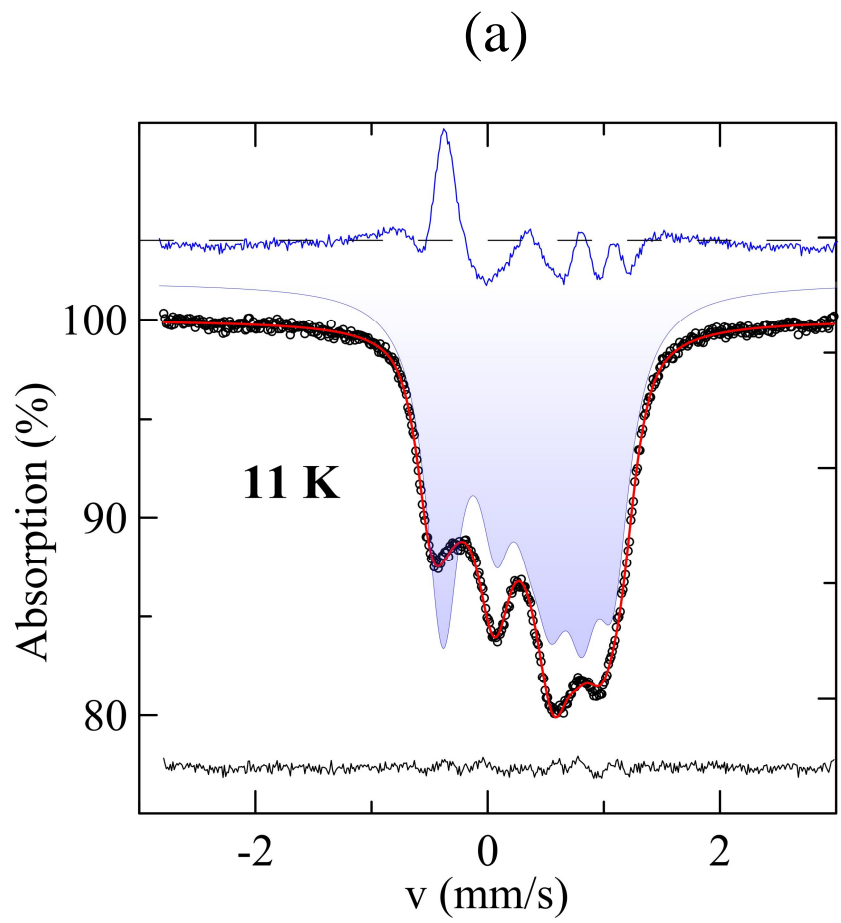

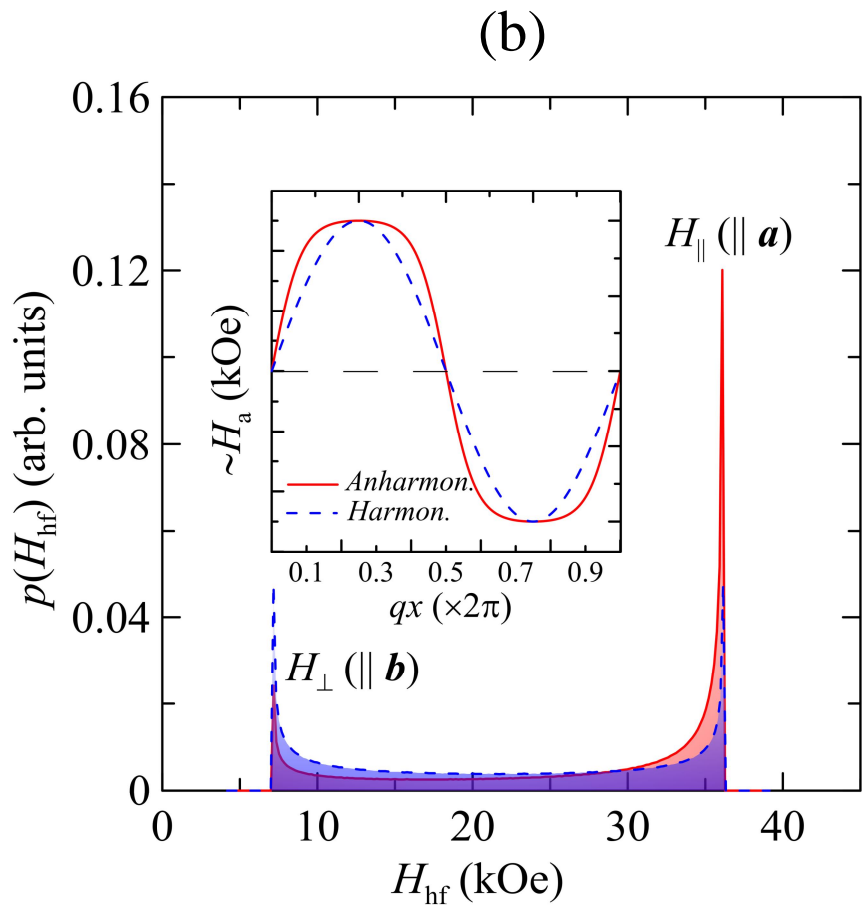

Fig. 3

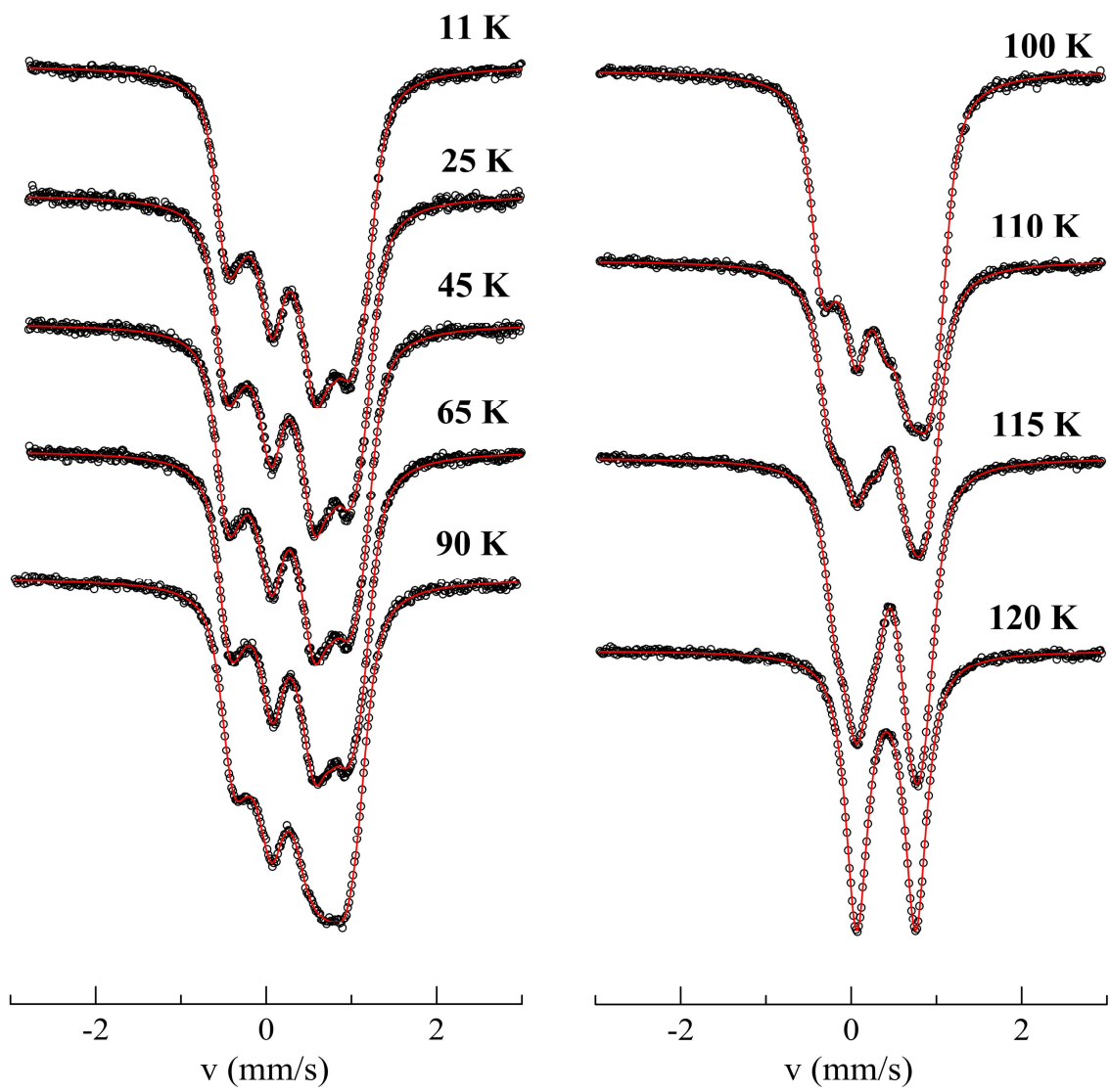

**Fig. 4**

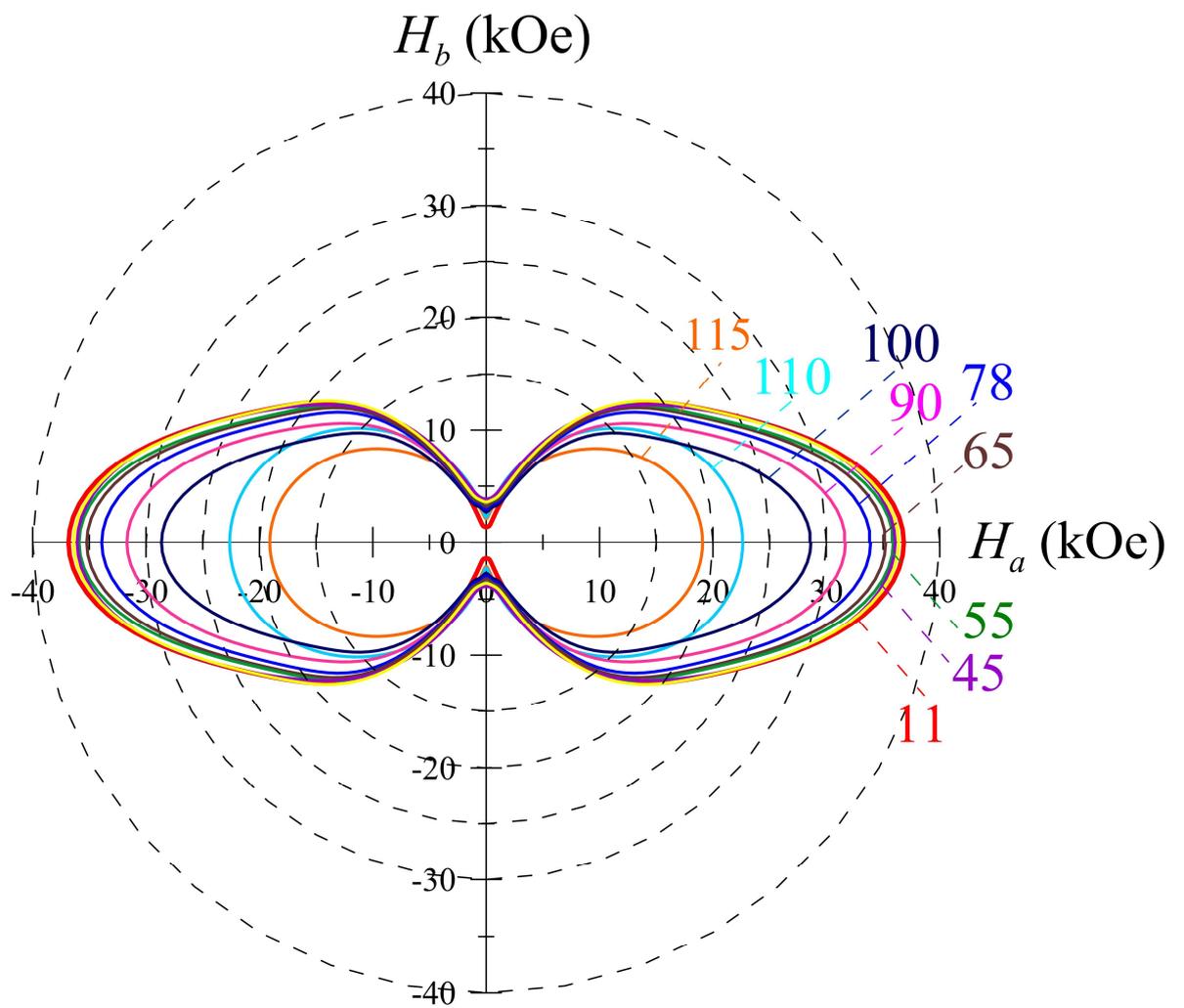

**Fig. 5**

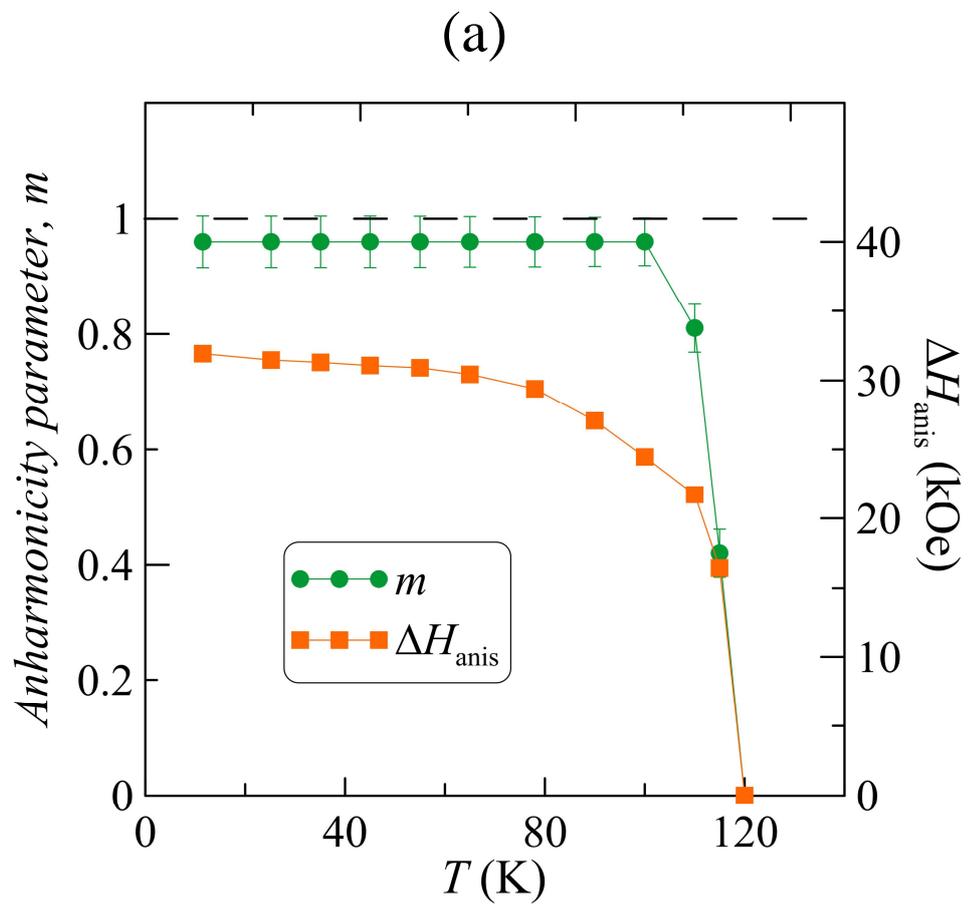

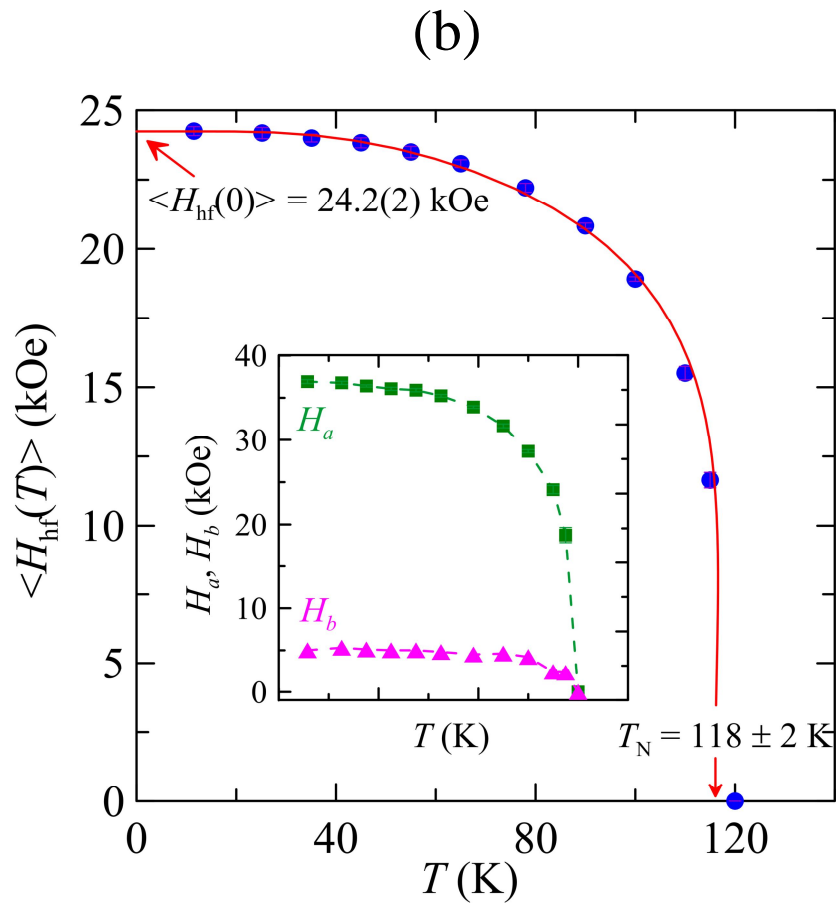

**Fig. 6**

(a)

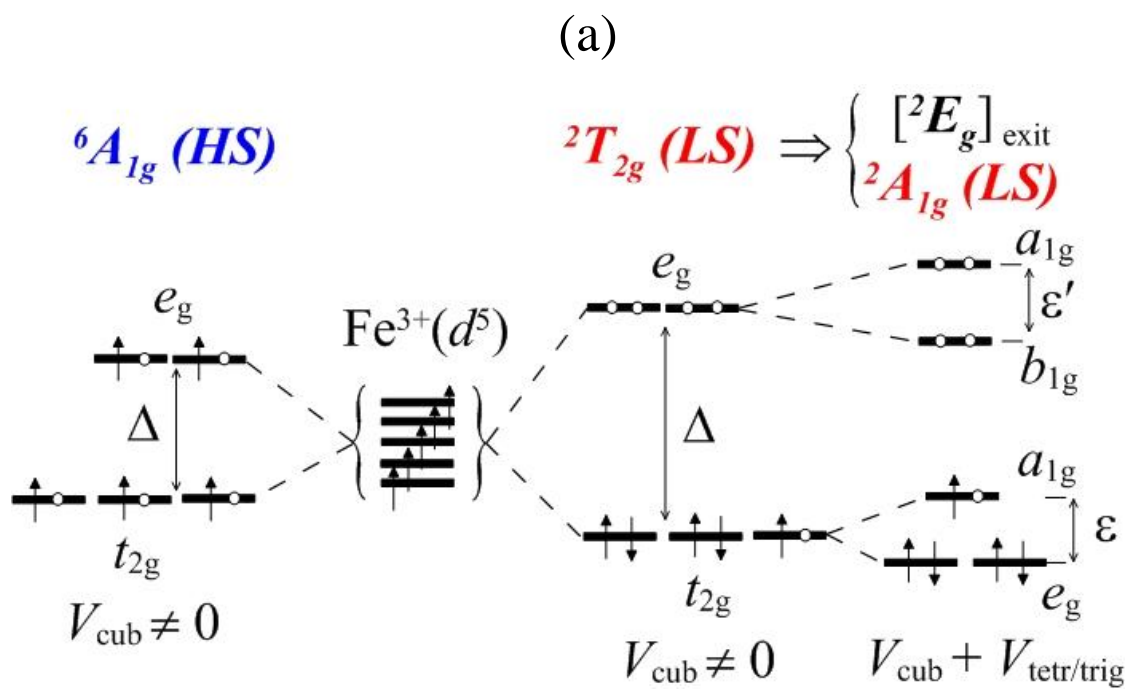

(b)

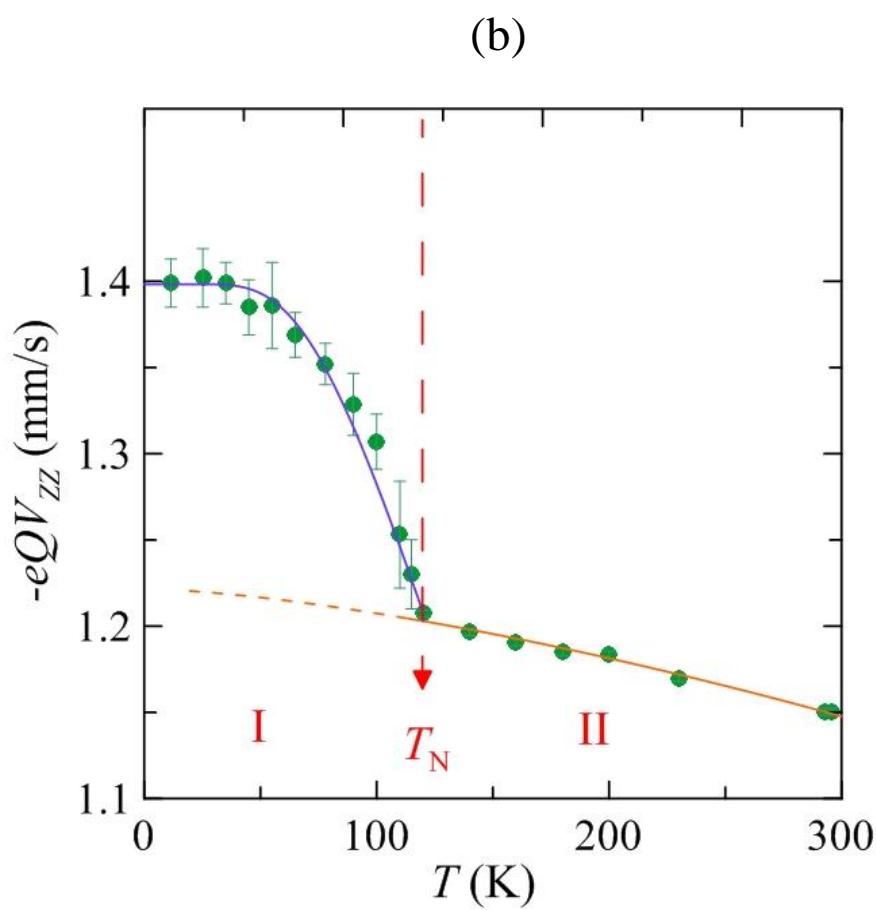

**Fig. 7**

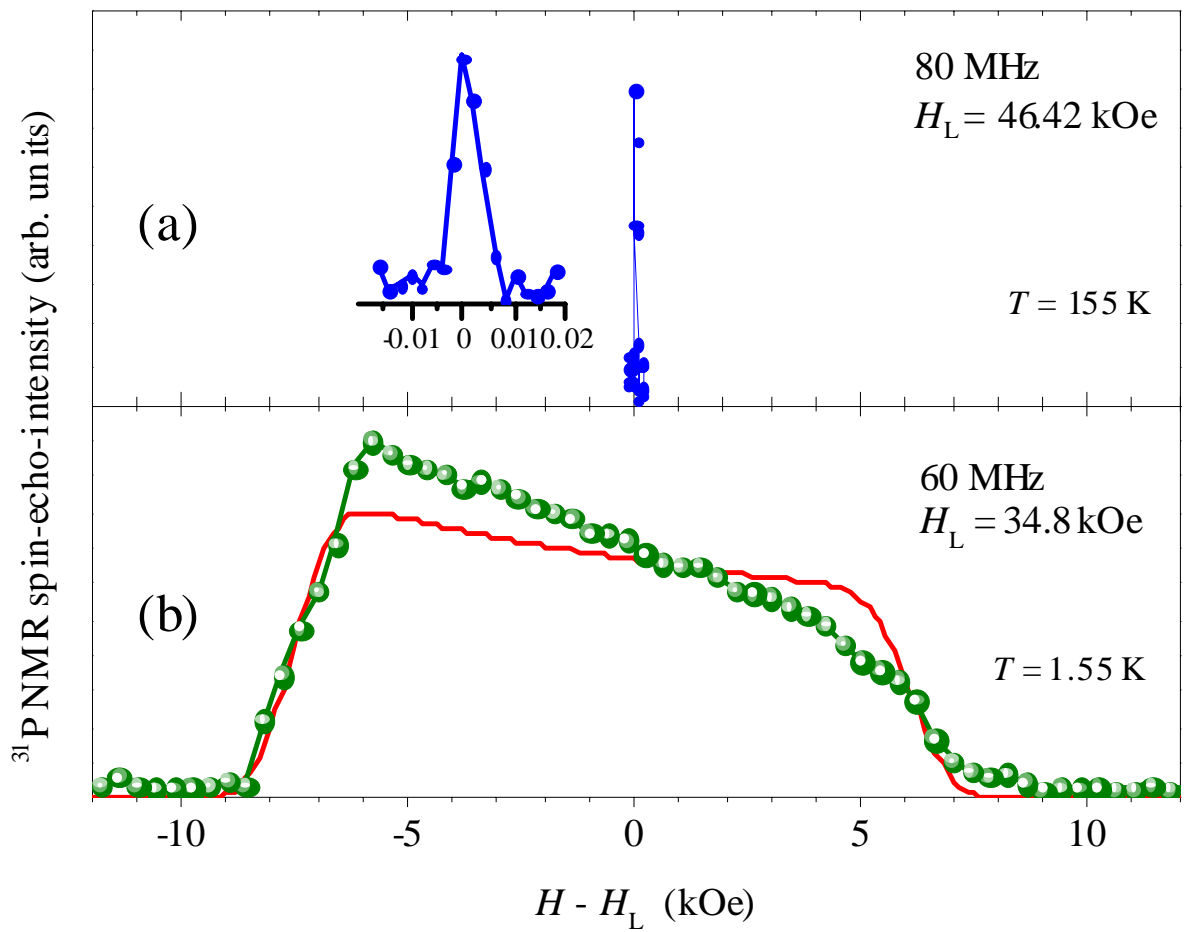

**Fig. 8**